# Wide range *Tα- Q$_α$* formula for real time application

*D. Ibadullayev, Yu.S.Tsyganov*

INP,050032, Almaty, Kazakhstan, JINR.

Ibadullayev@jinr.ru

**Abstract**

*A new wide-range formula $T_α=f(Q_α)$ to operate with the new DGFRS2 analog spectrometer installed at the DC-280 cyclotron facility is presented. The main goal of this formula application algorithm is to search the optimal time correlation recoil-alpha parameter directly during the execution of the C++ acquisition code. Note that the spectrometer operates together with the 48×128 strip DSSD (Double Side Strip Detector) detector and low-pressure pentane-filled gaseous detector. Comparison with others known formulas is performed for Z=119,120 nuclei.*

## 1. Introduction

With the discovery of the fission of uranium by Hann and Strassmann, the boundary of existence of nuclei was physically defined for the first time as a limit of stability of nuclei with spontaneous fission (SF) [1]. According to the theory [2] the fission barrier will rapidly decrease with growing Z (Z>92). In the macroscopic theory (liquid drop model) the situation with zero barrier occurs for the element with Z>100. The situation is changed in 1962 after observing the short SF half-life $T_{SF} ≈0.014$ s in $^{242}$Am known to have $T_{SF}> 3^{12}$ y in the ground state[3]. It means that nuclear structure does not disappear with increasing deformation but evolves and maintains an important role in nuclear fission process [4]. Elements with Z>100 were produced in the reactions induced by charged particles. In the beginning of present century new Z=114-118 elements were synthesized using the Dubna Gas-Filled Recoil Separator (DGFRS) [5-10]. That discovery confirmed the main role of shell effects in stability of superheavy nuclei. A key question relating to the discovery of new element and isotope is the probability $P_{err}$ that the event sequence observed is due to a random correlation of unrelated events. The magnitude of this probability allows readers and experimenters to judge the validity of the interpretation. Present work aimed to a consideration of wide-range formula to predict half-life time of new isotopes of superheavy elements. Namely, these estimates are used in "active correlation" method to provide a reasonable choice for *ER*(*E*vaporation *R*esidue)-α time interval to provide a radical suppression of background signals [11].

## 2. Formulae for prediction of time properties of superheavy nuclei

In [12] different formulae are presented to predict properties of alpha decaying nuclei. These formulae deal with $Z$, $Q_\alpha$ and $A$ as independent variables. From the other hand, in [13] the same predictions are made using only $Z$, $Q_\alpha$ parameters. In this paper parameter $d$ of the formula was varied to satisfy data presented in [5] for SHE as following: $T_\alpha^{calc}=10^{F(Z)}$, where $F(Z)=(aZ+b)\cdot Q^{-1/2}+c\cdot Z+d$. In [14] experimental and theoretical data for $Ac$ isotopes are presented, and the comparison of these results with one which estimated using above formula is shown in the Fig.1. These data are fitted well with optimal parameter $d=-27.5928$. In the *Table 1* an arbitrary choice of isotopes are presented with lower $Z$.

Table 1. d- Parameters for wide range of Z.

| Isotope | Z | $T_{1/2}$ | $Q_\alpha$, MeV | d |
|---|---|---|---|---|
| $^8$Be | 4 | 0.082 fs | 0.184 | 18.154 |
| $^{107}$Te | 52 | 3.1 ms | 4.0846 | -24.65124 |
| $^{108}$I | 53 | 36 ms | 4.25676 | -23.47449 |
| $^{110}$Xe | 54 | 93 ms | 4.0214 | -24.72388 |
| $^{144}$Nd | 60 | 2.29·10$^{15}$ y | 1.96088 | -23.14566 |
| $^{146}$Sm | 62 | 6.8·10$^7$ y | 2.600 | -24.31563 |
| $^{151}$Eu | 63 | 1.7·10$^{18}$ y | 2.01853 | -22.40255 |
| $^{148}$Gd | 64 | 71.1 y | 3.36209 | -25.0482 |
| $^{154}$Dy | 66 | 3·10$^6$ y | 3.02356 | -24.7324 |
| $^{152}$Ho | 67 | 2.7 min | 4.62916 | -26.42281 |
| $^{154}$Tm | 69 | 8.1 s | 5.2288 | -26.07 |
| $^{158}$Hf | 72 | 2.85 s | 5.54509 | -26.75256 |
| $^{164}$Os | 76 | 21 ms | 6.64097 | -26.74 |
| $^{176}$Hg | 80 | 20.3 ms | 7.0597 | -27.06 |
| $^{195}$At | 85 | 290 ms | 7.4927 | -26.553 |
| $^{216}$Rn | 86 | 45 mcs | 8.352 | -28.20686 |
| $^{207-209}$Ac | 89 (AverYang) | - | - | -26.72 |
| $^{236}$U | 92 | 2.34·10$^7$ y | 4.6517 | -27.99773 |
| $^{238}$U | 92 | 4.47·10$^7$ y | 4.3427 | -28.03537 |
| $^{241}$Am | 95 | 432.6 y | 5.76298 | -27.6253 |
| $^{252}$Fm | 100 | 1.06 d | 7.26808 | -27.9061 |
| $^{260}$Sg | 106 | 3.6 ms | 10.054 | -28.42474 |
| $^{279}$Ds | 110 (aver. [17]) | - | - | -28.318 |
| Fl | 114 (*aver.* [13]) | - | - | -28.0928 |

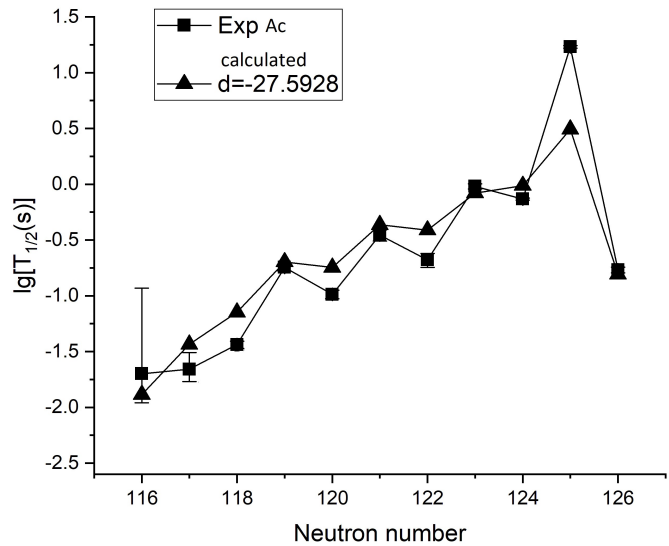

Fig.1 Measured [14] and calculated (*triangles*) dependences against half-life values of Ac isotopes.

Note, that in the positions of **Ac, Ds** and **Fl** the averaged values of **d**parameters are shown.

The wide range **d** parameter dependence against **Z** is shown in the Fig.2. Best fit is shown as power function ***d=-28.73672+74.1551·0.95105$^Z$***.

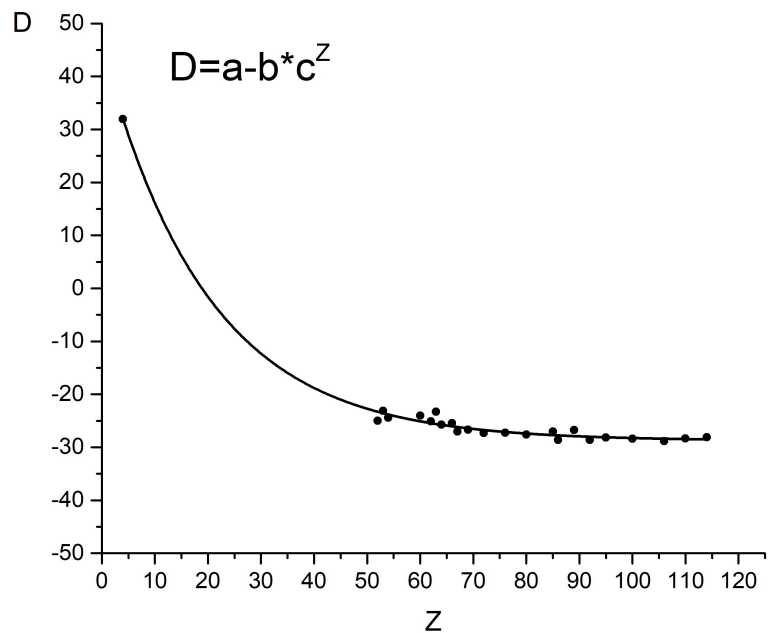

Fig.2 Dependence of *d*-parameter against *Z*. *(a=-28.73672; b= -74.1551; c=0.95105)*

## 3. Calculation of Z=119, 120 nuclei half life values

We calculated $T_\alpha$ values for the elements with **Z=119,120**. We plan to synthesize these elements at the **DGFRS2** setup [16] in a nearest future. $Q_\alpha$ values were taken from [13]. The results are presented in Table 2.

Table 2. $T_\alpha$ values calculated for **Z=119,120** nuclei.

| Z | $Q_\alpha$, MeV | $T_{1/2}$ (This paper) | $T_{1/2}$ [13] | $T_{1/2}$ [Royer] |
|---|---|---|---|---|
| 119 | 12.338 | 38 μs | 107.89 μs | 509.2 μs |
| 120 | 12.7 | 11 μs | 31.4 μs | 16.6 μs |

It should be noted, that $T_\alpha$ values calculated by Royer's formulae were performed for *A=294, 296* respectively. Known isotope properties were taken from [18].

## 4. Summary

New four parameter formula for $T\alpha=f(Q_\alpha)$ dependence is obtained. The $T_\alpha$ values were calculated for **Z=119,120** nuclei. Comparison with **Royer's** formulae as well as reported in [13] has been performed. We plan to apply these results as a base for application of "active correlations" method in the area of unknown isotopes synthesized in heavy ion induced complete fusion nuclear reactions at the **DGRFS2** setup. One general extra conclusion can be drawn here: namely, the described $T_\alpha=f(Q_\alpha)$ dependence can be used to predict reasonable value of unknown alpha decay parameters in the area with the lowest **Z** values.

Authors are indebted to A.Polyakov and A.Voinov for their help in data analysis.